\documentclass[reprint,pre,aps,  nolongbibliography
	footinbib
	]{revtex4-2}
 
\usepackage{geometry}
\usepackage{amsmath,amsthm,amssymb, mathtools,mathrsfs}
\usepackage{physics}
\usepackage{amsfonts, latexsym, amssymb}
\usepackage{bbm, dsfont}
\usepackage{caption}
\usepackage{graphicx}
\usepackage{float}
\usepackage{url}
\usepackage{xcolor}
\usepackage{hyperref}
\usepackage{tikz}
\usetikzlibrary{decorations.pathmorphing,cd,decorations.markings,calc,fadings}
\hypersetup{
    colorlinks=true,
    linkcolor=blue,
    filecolor=magenta,      
    urlcolor=cyan,
    citecolor=magenta,
}
\usepackage{enumerate}
\usepackage{bm}

\definecolor{darkgreen}{rgb}{0.1,0.6,0.1}

\newcommand{\be}{\begin{equation}} \newcommand{\ee}{\end{equation}}
\newcommand{\ben}{\begin{equation*}} \newcommand{\een}{\end{equation*}}

\newcommand{\bea}{\begin{equation} \begin{aligned}} \newcommand{\eea}{\end{aligned} \end{equation}}

\def\repa{\raise4pt\hbox{$\square$}\mkern-14mu\raise-4pt\hbox{$\square$}}
\def\repab{\overline{\raise4pt\hbox{$\square$}\mkern-14mu\raise-4pt\hbox{$\square$}\mkern-1mu}}

\usepackage{tikz}
\usetikzlibrary{shapes.misc}
\tikzset{cross/.style={cross out, draw=black, ultra thick, minimum size=3*(#1-\pgflinewidth), inner sep=0pt, outer sep=0pt},
cross/.default={5pt}}

\begin{document}

\title{String Duals of Two-Dimensional Yang-Mills and Symmetric Product Orbifolds
}
\author{Shota Komatsu$^{a}$ and Pronobesh Maity$^{b}$}
\affiliation{${}^{a}$Department of Theoretical Physics, CERN, 1211 Meyrin, Switzerland}
\affiliation{${}^{b}$Laboratory for Theoretical Fundamental Physics, EPFL, Rte de la Sorge, Lausanne, Switzerland}
\email{shota.komatsu@cern.ch}
\email{pronobesh.maity@epfl.ch}

\date{\today}

\begin{abstract}
    We propose a bosonic string dual to large $N$ chiral  Yang-Mills in two dimensions at finite 't Hooft coupling. The worldsheet theory is a $\beta$-$\gamma$ system deformed by a chiral  Polchinski-Strominger term. 
    We reproduce the partition function on a torus, cylinder three-point amplitudes, and the area law for Wilson loops. We also present candidate string duals to symmetric product orbifolds for general seed CFTs with $c<24$ and their $T\bar{T}$-, $J\bar{T}$-deformations. The results hint at interrelations among confining, nonrelativistic and matrix strings, and AdS$_3/$CFT$_2$.
\end{abstract}

\maketitle
\section{Introduction}
The idea of reformulating gauge theory in terms of strings has a long history, originating with ’t Hooft’s large $N$ expansion \cite{tHooft:1973alw} and loop equations \cite{Makeenko:1979pb,Makeenko:1980vm}. A major early success came from {\it non-critical string theory} \cite{Gross:1989vs,Brezin:1990rb,Douglas:1989ve}; dualities between simple large $N$ matrix models and string theory in less than one dimension. The idea later culminated in the AdS/CFT correspondence \cite{Maldacena:1997re}, offering concrete examples of dualities between supersymmetric gauge theories and string theory in Anti-de Sitter (AdS) spacetime.

A common feature in these dualities is the emergence of an extra spatial direction---often called “holographic” or “Liouville”---linked to an additional scalar mode on the string worldsheet. 
While this is often viewed as a hallmark of gauge/string duality, recent lattice studies of confining fluxtubes \cite{Teper:2009uf,Athenodorou:2010cs,Athenodorou:2011rx,Dubovsky:2013gi,Dubovsky:2014fma} challenge this paradigm, suggesting that the extra scalar is absent for Yang-Mills theory in three and four dimensions.
Relatedly, effective field theories for confining fluxtubes \cite{Polchinski:1991ax,Aharony:2013ipa,Hellerman:2014cba} are formulated without an extra scalar, although they are not UV-complete. This leads to a question:\\
{\it Is there a UV-complete string dual of a confining gauge theory without an extra scalar?}

In this work, we answer yes, presenting a dual for the simplest confining gauge theory; chiral 2d Yang-Mills (YM). As a byproduct, we also propose candidate duals to symmetric product orbifolds for any seed CFTs with $c<24$ and their $T\bar{T}$- and $J\bar{T}$-deformations. Notably, none of these constructions require supersymmetry, unlike many top-down examples of holography.

\section{String dual of Chiral 2d YM}
2d YM has been a subject of interest due to its confining nature and solvability. Importantly, its partition function at large $N$ factorizes into chiral and anti-chiral parts, each of which receives $1/N$ corrections. In addition, there are non-factorizable corrections, $Z_{\text{2d YM}}=Z_{\rm chiral} Z_{\rm anti}+\#/N^2+ \cdots$. 

Key developments in the past include
\begin{itemize}
\item \cite{Gross:1992tu,Gross:1993hu} showed that its $1/N$ expansion can be interpreted as a genus expansion of string theory, although an explicit worldsheet theory reproducing the expansion was not constructed. 
\item \cite{Cordes:1994sd} identified a dual of the zero-coupling limit of 2d YM as topological string. However, it remains to be understood how to generalize it to finite coupling. See \cite{Benizri:2025xmz} for a recent proposal on the deformation of Gromov-Witten theory.
\item A ``rigid topological string" action was proposed as a candidate dual to 2d YM at finite coupling \cite{Horava:1995ic}. Due to the unconventional structure of the worldsheet theory, however, its analysis poses technical challenges. This approach has recently been revisited, with detailed studies focusing on the zero-coupling regime \cite{Aharony:2023tam}.
\item It was shown that topological string on certain Calabi-Yau manifolds reproduce the partition function of the $q$-deformed 2d YM at finite coupling \cite{Vafa:2004qa,Aganagic:2004js}. See also \cite{Caporaso:2005ta, Caporaso:2005fp, Caporaso:2006kk} for further analyses including the undeformed limit.
\end{itemize}
Below we present a bosonic string dual of chiral 2d YM that allows computation of observables beyond the partition function, including scattering amplitudes---unlike earlier proposals. It however requires a {\it flat} target space: $\mathbb{R}^{1,1}$, torus, or cylinder.
\subsection{Worldsheet action}
Our proposed dual for $U(N)$ 2d YM at large $N$ with 't Hooft coupling $\lambda =g_{\rm YM}^2N$ is
\begin{align}\label{eq:actionYM}
S=&\int \frac{d^2z}{2\pi} \Big[ \beta \bar{\partial}X^{+}+\bar{\beta}\partial X^{-}+\frac{q}{2}\mathcal{L}_{CLD}  \nonumber\\
& +\frac{\lambda\pi}{2}\left(\partial X^{+} \bar{\partial} X^{-}-\bar{\partial}X^{+}\partial X^{-}\right)\Big]+ \text{bc-ghost}\,,
\end{align}
where $X^{\pm}$ are light-cone coordinates of the target space $X^{\pm}=X^{1}\pm X^{0}$ and $\mathcal{L}_{CLD}$ is a {\it chiral composite linear dilaton} action,
\begin{align}\label{eq:CCLDaction}
    \mathcal{L}_{CLD}=2\,\partial \varphi\,\bar{\partial}\varphi+\hat{R}\,\varphi\,, \quad \varphi =\log (\frac{\partial X^{+}\bar{\partial}X^{-}}{R^2}) \,,
\end{align}
where $R$ is the length-scale of the target space and $\hat{R}$ is the worldsheet Ricci scalar. A few comments are in order
\begin{itemize}
\item The anomaly cancellation sets $q=1$ (see below).
\item The $\beta$-$\gamma$ system shows up also in nonrelativistic string theory \cite{Gomis:2000bd,Danielsson:2000gi,Danielsson:2000mu}. Our theory can be viewed as its non-critical generalization.
\item $\mathcal{L}_{CLD}$ resembles the Polchinski-Strominger term  \cite{Polchinski:1991ax} in the Polyakov formalism \cite{Hellerman:2014cba}, differing only by the lack of $\bar{\partial}X^{+}\partial X^{-}$ in $\varphi$. 
\item Despite high nonlinearity due to $\mathcal{L}_{CLD}$, the path integral can be evaluated explicitly by first integrating out $\beta$'s, which sets $X^{+}$ ($X^{-}$) to be (anti-)holomorphic.
\item Similar actions were proposed in the studies of lightcone-gauge string field theory \cite{Baba:2009ns}, Green-Schwarz string in the semi-lightcone gauge \cite{Kazama:2010ys} and $T\bar{T}$-deformation \cite{Callebaut:2019omt}. We discuss the relationships in the Supplemental Material and note that the last one does not exhibit conformal invariance in its current form.
\end{itemize}
From the action \eqref{eq:actionYM}, the operator product expansion (OPE) can be derived \cite{toappear}:
\begin{align}\label{eq:OPE}
&\beta(z)\, X^{+}(w)\sim -\frac{1}{z-w},\quad \bar{\beta}(\bar{z})\,X^{-}(\bar{w})\sim-\frac{1}{\bar{z}-\bar{w}}\\
&\beta(z)\,\beta(w)\sim 2q \, \partial_z\partial_w \left[\frac{1}{(z-w)^2}\,\frac{1}{\partial_z X^{+}(z)\,\partial_w X^{+}(w)}\right]\,,\nonumber
\end{align}
The stress tensor of the matter sector reads
\begin{align}\label{eq:stress}
T(z)&=-\beta \partial X^{+}(z)+2q \{X^{+},z\} \, ,
\end{align}
with $\{X^{+} ,z\}$ being the Schwarzian derivative. The OPE of stress tensors can be computed from \eqref{eq:OPE} as
\begin{equation}\label{TT_OPE}
	T(z)\,T(w)=\frac{1+12q}{(z-w)^4}+\frac{2T(w)}{(z-w)^2}+\frac{\partial_w T(w)}{z-w}\,.
\end{equation}
This shows that the matter action in \eqref{eq:actionYM} has central charge $c=2+24q$ and $q=1$ is needed for the conformal anomaly cancellation.

\subsection{Torus partition function}
Let us take the target space to be Euclidean torus $\mathbb{T}^{2}_{\rm TS}$ with modulus $\zeta$ and size $R$,
\begin{align}
\begin{aligned}
&X^{+}\to X=X^{1}+iX^{2}\,,\quad X^{-} \to \bar{X}=X^{1}-iX^{2}\,,\\
&(X\sim X+2\pi R\sim X+2\pi R \zeta) \, ,
\end{aligned}
\end{align}
and study the genus-1 worldsheet partition function $\mathcal{Z}_{\rm ws}(R,\zeta)$ on this background. Below we sketch the computation relegating the details to Supplemental Material.

First, integrating out oscillator modes of $\beta$ cancels the $bc$-ghost contributions and sets $X$ ($\bar{X}$) to be (anti-)holomorphic except for zero modes. This forces $X$ to be a map from the worldsheet torus $\mathbb{T}_{\rm ws}^2$ to $\mathbb{T}_{\rm TS}^2$, characterized by four winding numbers, $m^{1,2}$ and $w^{1,2}$:
\begin{align}
\frac{X(z,\bar{z})}{R} = \frac{(M-\bar{\tau}W)z+(\tau W-M)\bar{z}}{\tau-\bar{\tau}}\,.
\end{align}
Here $M=m^{1}+\zeta m^{2}$ and $W=w^{1}+\zeta w^{2}$, and $z$ parametrizes $\mathbb{T}^{2}_{\rm ws}$ as $z \sim z+2\pi \sim z+2\pi \tau$.

Second, the integral of zero modes of $\beta$'s produces the delta function that localizes the worldsheet to {\it covering maps} from $\mathbb{T}_{\rm WS}^2$ to $\mathbb{T}_{\rm TS}^2$, leading to the following expression for $\mathcal{Z}_{\rm WS}$,
\begin{align}\label{eq:moduli2dYM}
\mathcal{Z}_{\rm WS}=\!\!\sum_{\{m^{a},w^{a}\}}\frac{\zeta_2}{|W|^2}\int_{\mathcal{F}}\frac{d^2\tau}{\tau_2}\delta^{(2)}\left(\tau-\dfrac{M}{W}\right)e^{-S_{m,w}}\,,
\end{align}
where $\zeta_2={\rm Im}\zeta$, $\tau_2={\rm Im}\tau$ and $\mathcal{F}$ is the fundamental domain of the torus moduli and $S_{m,w}$ is an on-shell action,
\begin{align}
S_{m,w}=\frac{\lambda\pi^2R^2}{2\tau_2}|\bar{W}\tau-\bar{M}|^2\,.
\end{align}
To evaluate the moduli integral \eqref{eq:moduli2dYM}, we extend the integration domain to the upper half plane using part of the sum over $\{m^{a},w^{a}\}$ \cite{Polchinski:1985zf}, resulting in
\begin{align}\label{eq:2dYMWS}
\mathcal{Z}_{\rm WS}=\sum_{K=1}^{\infty}\frac{e^{-\frac{K\lambda  A}{2}}}{2K}\sum_{\substack{ad=K\\a,d\in \mathbb{Z}_{+}}}(a+d)\,,
\end{align}
where $A$ is the area of $\mathbb{T}^{2}_{\rm TS}$.
This precisely matches the free energy of chiral 2d YM at large $N$ \cite{Witten:1991we,Gross:1993hu,Douglas:1989ve}. Note that the zero winding sector $K=0$ does not contribute thanks to the CLD action as will be explained around \eqref{eq:zerowidiingsingular}.

\subsection{Three-point amplitudes}
Due to the absence of local dynamics, wave functions of chiral 2d YM on $S^1$ are given simply by class functions of the holonomy $U$ \cite{Witten:1991we}. In particular, a convenient basis at large $N$ is the multi-trace basis $\{|\sigma\rangle\}$,
\begin{align}
		\Upsilon_{\sigma}(U)= \langle U | \sigma\rangle =\prod_{\ell} \left(\text{tr} (U^\ell)\right)^{k_{\ell}}\,,
	\end{align}
where $k_{\ell}$ denotes the number of cycles of length $\ell$ in $\sigma \in S_{r}$ with $r=\sum_{\ell} \ell k_{\ell}$. As discussed in \cite{Minahan:1993np}, $|\sigma\rangle$ corresponds to a multi-string state with $k_{\ell}$ strings with winding number $\ell$. It can be expressed with string creation operators as 
\begin{align}
|\sigma\rangle =\prod_{\ell}(a_{\ell}^{\dagger})^{k_{\ell}}|\Omega\rangle\,,
\end{align}
where $[a_{k},a_{l}^{\dagger} ]=k\delta_{k,l}$. The Hamiltonian on $S^1$ of radius $R$ is then given by \cite{Minahan:1993np}
\begin{align}
\begin{aligned}
&H=H_{\rm free}+H_{\rm int}\,, \qquad H_{\rm free}=\pi \lambda R \sum_{m}a_{m}^{\dagger}a_m \,,\\
&H_{\rm int}=\frac{\pi R \lambda}{N}\sum_{n,m}a_{m+n}^{\dagger}a_m a_n +a_m^{\dagger}a_n^{\dagger}a_{m+n}\,.
\end{aligned}
\end{align}
Viewing $H_{\rm int}$ as the interaction Hamiltonian, one can compute the perturbative S-matrix of winding strings. In particular, the $2\to 1$ amplitude reads
\begin{align}\label{eq:3ptYM}
\mathcal{S}=-i\frac{4\lambda\pi^2 R}{N}w_1w_2w_3\delta(E_1+E_2-E_3)\,\delta_{w_1+w_2,w_3}\,,
\end{align}
where $w_k$  and $E_k=\lambda \pi Rw_k$ are the winding number and the energy of the $k$-th string respectively.

We now reproduce it from genus-$0$ amplitudes in string theory \eqref{eq:actionYM} on a Lorentzian cylinder $X^{\pm}\sim X^{\pm}+2\pi R$. The vertex operator for the $w_k$ winding state is \cite{toappear, Gomis:2000bd}
\begin{align}
\mathcal{V}_k (z_k)=e^{iw_kR\int^{z_k} (\beta dz- \bar{\beta}d\bar{z})}   e^{-i E_kX^{0}}\,.
\end{align}
Inserting them in the path integral and integrating out $\beta$'s, we find that $(X^{+},X^{-})$ localizes to the Maldelstam map $(\rho,\bar{\rho})$,
\begin{align}
X^{+}(z)\mapsto \rho(z)=-i\sum_{k}w_k R\log (z-z_k)\,.
\end{align}
As a result, we obtain the following expression for the three-point amplitude
\begin{align}
&G_3=g_s\langle\prod_{k=1}^{3}\mathcal{V}_k(z_k) \rangle\\
&=ig_s \mathcal{N}_{S^2}\,2\pi\,\delta(\sum_{k}E_k)\,2\pi R\,\delta_{\sum_kw_k}\,
 e^{-\Gamma [\rho]}\prod_{i<j}|z_{ij}|^2\,,\nonumber
\end{align}
where $\mathcal{N}_{S^2}$ is the normalization coming from the sphere partition function and the path-integral measure and $\Gamma[\rho]$ is the renormalized on-shell action for $\frac{q}{2}\int \mathcal{L}_{\rm CLD}$ \cite{Baba:2009ns,Mandelstam:1973jk},
\begin{align}
e^{-\Gamma[\rho]}=\prod_{k=1}^{n} (w_k)^{2q}\left|\sum_{k=1}^{n}w_kz_k\right|^{-4q}\left|\prod_{I=1}^{n-2}\frac{\partial^2\rho(Z_I)}{R}\right|^{q}\,,
\end{align}
where $Z_I$'s are the ``interaction points", where $\partial\rho$ vanishes. Here we absorbed the divergence around the interaction points by renormalizing $g_s$, as in the lightcone string field theory \cite{Baba:2009ns,Mandelstam:1973jk,Green:1987mn}. Setting $n=3$ and $q=1$ gives $e^{-\Gamma}=w_1w_2w_3 \prod_{i<j}|z_{ij}|^{-2}$. We thus have
\begin{align}\label{eq:3ptstring}
G_3=i(2\pi)^2 R g_s\,\mathcal{N}_{S_2}  w_1w_2w_3 \,\delta(\sum_kE_k) \,\delta_{\sum_k w_k} \,,
\end{align}
Continuing the third state from the in- to out-state by $E_i\to -E_i$ and $w_i\to -w_i$, \eqref{eq:3ptstring} reproduces \eqref{eq:3ptYM} under the identification of parameters \footnote{Precisely speaking, the dictionary between parameters has ambiguity of arbitrary rescaling, $g_s \to r g_s$ and $\mathcal{N}_{S^2}\to r^2 \mathcal{N}_{S^2}$.},
\begin{align}
g_s=\frac{\lambda R^2}{2N}\,,\quad \mathcal{N}_{S^2}=\frac{2}{R^2}\,.  
\end{align}

\subsection{Wilson loop}
We now consider a disk amplitude for the theory on $\mathbb{R}^2$. It is conveneint for the analysis to take the ``winding gauge" \cite{McGough:2016lol}
\begin{align}\label{eq:windinggauge}
\partial X=\bar{\partial}\bar{X}=1\,.
\end{align}
In this gauge, the disk amplitude of a string ending on a curve $\mathcal{C}\subset \mathbb{R}^2$ is given by evaluating the on-shell action. The result reads
\begin{align}
\mathcal{Z}_{D^2} \propto e^{-\frac{\lambda A}{2}}\,,
\end{align}
reproducing the area law for the Wilson loop in 2d YM. 

\section{Symmetric Product Orbifolds}
\subsection{Action and partition function}
We next provide evidence that a generalization of \eqref{eq:actionYM} is a potential dual to symmetric product orbifolds for arbitrary seed CFTs. The worldsheet action reads
\begin{align}\label{eq:symproaction}
\begin{aligned}
S_{\rm sym}=&\int \frac{d^2z}{2\pi} \left[ \beta \bar{\partial}X+\bar{\beta}\partial \bar{X}+\frac{q}{2}\mathcal{L}_{CLD}\right]+S_{\rm seed} \\
& -B\int \frac{d^2 z}{4\pi} \left(\partial X \bar{\partial} \bar{X}-\bar{\partial}X\partial \bar{X}\right) + \text{bc-ghost}\,,
\end{aligned}
\end{align}
with $q=(24-c)/24$, where $c$ and $S_{\rm seed}$ are the central charge and the action of the seed CFT.

The genus-1 partition function on a torus ($X\sim X+2\pi R\sim X+2\pi R \zeta $) can be evaluated as in 2d YM thanks to the localization,
\begin{align}\label{eq:symproductWS}
\mathcal{Z}_{\rm WS}=\sum_{a,d=1}^{\infty}\frac{p^{ad}}{ad}\sum_{b=0}^{d-1}\mathcal{Z}_{\rm seed}\left(\frac{b+a\zeta}{d},\frac{b+a\bar{\zeta}}{d}\right)\,,
\end{align}
where $p=e^{B\,\text{Area}/(2\pi)}$ and $\mathcal{Z}_{\rm seed}$ is the torus partition function of the seed CFT. This agrees precisely with the grand-canonical free energy of the symmetric product orbifold \cite{Dijkgraaf:1996xw}. Note that the duality relates string theory to a grand-canonical ensemble of CFTs rather than a theory with a fixed central charge, as was the case with supersymmetric tensionless AdS$_3$/CFT$_2$ \cite{Eberhardt:2020bgq, Aharony:2024fid}. 
\subsection{Holomorphic map vs. covering map} Let us briefly discuss higher genus corrections. Due to the $\beta$ integrals, the string path integral always localizes to holomorphic maps $\rho$. However, holomorphic maps in general admit extra branch points ($\partial\rho (Z_I)=0$) while the partition function of symmetric product orbifolds is given by a sum over {\it covering maps} \cite{Lunin:2000yv}; holomorphic maps without such branch points.

This discrepancy can be resolved for seed CFTs with $c<24$. Evaluating the on-shell action of $\mathcal{L}_{\rm CLD}$ around the branch point $Z_I$, we get
\begin{align}\label{eq:zerowidiingsingular}
\exp\left[-\frac{q}{2}\int \frac{d^2z}{2\pi}\mathcal{L}_{\rm CLD}\right] \supset  |2\epsilon_I\partial^2 \rho (Z_I) |^{q}\, ,
\end{align}
where $\epsilon_I$ is a radius of a small circle around $Z_I$ excised for the regularization. This shows that, in the limit $\epsilon_I \to 0$,  the maps with branch points do not contribute as long as $q>0$, i.e.~$c<24$. The same mechanism removes the zero-winding sector in the partition functions \eqref{eq:2dYMWS} and \eqref{eq:symproductWS}.
\subsection{Comparison to AdS$_3$/CFT$_2$}The bosonic action for string on AdS$_3$ supported by $k$ units of NSNS flux reads \cite{Giveon:1998ns}
\begin{align}
\begin{aligned}\label{eq:AdS3action}
&S_{\text{AdS}_3}=\\
&k\int \frac{d^2 z}{4\pi}(4\partial \Phi\bar{\partial}\Phi+\beta\bar{\partial}\gamma+\bar{\beta}\partial \bar{\gamma}-e^{-2\Phi}\beta\bar{\beta}-\frac{\hat{R}\Phi}{k})\,.
\end{aligned}
\end{align}
For $k=1$, it was argued in \cite{Eberhardt:2019ywk} that the path integral localizes to classical solutions near the boundary given by $\Phi= -\log \epsilon -\frac{1}{2}\log \partial \gamma \bar{\partial} \bar{\gamma}$. After this substitution and appropriate rescalings, the action \eqref{eq:AdS3action} coincides with our proposed action \eqref{eq:symproaction}. Thus, our action can be viewed as a theory where the (non-dynamical) Liouville mode $\Phi$ is integrated out. A key difference is that our action \eqref{eq:symproaction} always exhibits localization and produces the correct partition function for any seed CFT with $c<24$ while the result of \cite{Eberhardt:2019ywk} holds only for specific supersymmetric seed CFTs.
\section{$T\bar{T}$ and $J\bar{T}$ deformations}
We now further extend our action \eqref{eq:symproaction} to describe symmetric product orbifolds of $T\bar{T}$- and $J\bar{T}$-deformed CFTs. Technical details are explained in the Supplemental Material.
\subsection{$T\bar{T}$-deformation} We conjecture that symmetric product orbifolds of $T\bar{T}$-deformed CFT \cite{Smirnov:2016lqw} is dual to the following action
\begin{align}\label{eq:TTbaraction}
S_{T\bar{T}}=S_{\rm sym}-\frac{\mu}{\pi}\int\frac{d^2z}{2\pi}\beta\bar{\beta}\,,
\end{align}
where $S_{\rm sym}$ is given by \eqref{eq:symproaction}. The simplest way to understand the connection to the $T\bar{T}$-deformation is to employ the winding gauge \eqref{eq:windinggauge}. In this gauge, the worldsheet stress tensor for $S_{\rm sym}$ is
\begin{align}\label{eq:virasoro}
T(z)=-\beta \partial X+2q \{X,z\}+T_{\rm seed}\,\to\, -\beta+T_{\rm seed}\,,
\end{align}
where $T_{\rm seed}$ is the stress tensor of the seed CFT. Imposing the Virasoro constraint $T(z)=0$, we get $\beta=T_{\rm seed}$. Thus the deformation term in \eqref{eq:TTbaraction} can be identified with the operator $T_{\rm seed}\bar{T}_{\rm seed}$. 
\begin{itemize}
\item A related observation connecting the deformation of nonrelativistic string and $T\bar{T}$-deformations has been made in \cite{Blair:2020ops,Blair:2024aqz,Bergshoeff:2025uut}.
\item Similar actions were obtained by expanding deformed AdS$_3$ actions near the AdS boundary \cite{Chakraborty:2019mdf,Hashimoto:2019wct}. A key difference is that the boundary expansion in \cite{Chakraborty:2019mdf,Hashimoto:2019wct} comes with an extra Liouville mode and hence the dual CFT inevitably includes the linear dilaton as part of the seed CFT. On the other hand, our action can be defined for more general seed CFT. Another important point is that the full deformed AdS$_3$ action before the boundary expansion is not dual to the symmetric product orbifold (cf.~\cite{Eberhardt:2021vsx}) and the boundary expansion only captures a subsector of the full theory in a certain limit.
\item In the special case of the symmetric product orbifold of $T\bar{T}$-deformed theory with 4 free bosons and 4 free fermions, \cite{Dei:2024sct} analyzed the corresponding worldsheet dual using tensionless AdS$_3$/CFT$_2$. They reproduced the partition function and identified the connection between the worldsheet deformation and the $T\bar{T}$-operator.
\end{itemize}

The relation to the $T\bar{T}$ deformation can be verified more directly by computing the torus partition function. To do so, we integrate out $\beta$ and $\bar{\beta}$ to obtain a Polyakov-type action
\begin{align}
S_{T\bar{T}} =\int \frac{d^2z}{2\mu} \bar{\partial} X \partial \bar{X}+\cdots \, ,
\end{align}
and use the winding gauge $\partial X=\bar{\partial}\bar{X}=1$; see Supplemental Material for details. The result reads
\begin{align}\label{eq:TTbardefresult}
\mathcal{Z}_{\rm WS}=\sum_{K> 0} p^{K}\mathcal{T}_{K}[\mathcal{Z}_{+}]+p^{-K}\mathcal{T}_{K}[\mathcal{Z}_{-}]\,,
\end{align}
where $\mathcal{T}_K$ is the Hecke operator \cite{Dijkgraaf:1996xw},
\begin{align}\label{eq:Hecke}
\mathcal{T}_K[f]=\frac{1}{K}\sum_{\substack{ad =K\\0\leq b<d, \,d>0}}f\left(dR, \frac{a\zeta+b}{d},\frac{a\bar{\zeta}+b}{d}\right)\,,
\end{align}
and $\mathcal{Z}_{\pm}$ are the partition functions with winding number $\pm 1$,
\begin{align}\label{eq:defZ1}
\mathcal{Z}_{\pm}=\frac{\pi R^2\zeta_2}{ \mu}\int_{\mathcal{H}_{+}}\frac{d^2\tau}{(\tau_2)^2} e^{-\frac{\pi^2R^2}{\mu \tau_2}|\zeta\mp\tau|^2}\mathcal{Z}_{\rm seed}(\tau,\bar{\tau})\,,
\end{align}
where $\mathcal{H}_{+}$ are the upper-half-plane. Crucial differences from symmetric product orbifolds \eqref{eq:symproductWS} are the contribution from negative winding sectors and the absence of localization of the worldsheet moduli.

Expanding $\mathcal{Z}_{\rm seed}$ in   \eqref{eq:defZ1}, $\mathcal{Z}_{+}$ reproduces the standard $T\bar{T}$-deformed spectrum \cite{Smirnov:2016lqw},
\begin{align}
\begin{aligned}
&E_{(+)}(\mu)=\frac{\pi R}{\mu}\left(-1+\sqrt{1+\frac{\mu^2P^2}{\pi^2R^2}+\frac{2 \mu E}{\pi R}}\right)\,,\\ &P_{(+)}(\mu)=P\,,
\end{aligned}
\end{align}
where $E$ and $P$ are undeformed energy and momentum. On the other hand, the expansion of $\mathcal{Z}_{-}$ gives
\begin{align}
\begin{aligned}
&E_{(-)}(\mu)=\frac{\pi R}{\mu}\left(1+\sqrt{1+\frac{\mu^2P^2}{\pi^2 R^2}+\frac{2 \mu E}{\pi R}}\right)\,,\\ &P_{(-)}(\mu)=P\,.
\end{aligned}
\end{align}
In the undeformed limit $\mu\to 0$, $E_{(-)}$ diverges as $\propto 1/\mu$. Thus, negative winding sectors can be viewed as non-perturbative corrections to the full partition function, as was noted also in \cite{Benjamin:2023nts}.
\subsection{$J\bar{T}$-deformation} The discussion around \eqref{eq:virasoro} suggests  that the $J\bar{T}$-deformation \cite{Guica:2017lia} corresponds to deforming the worldsheet action by $\int d^2 z \,\bar{\beta} J $. For concreteness, let us assume that the seed CFT contains a compact boson of radius $r$
\begin{align}
S_{\mathbb{S}^1}=\int \frac{d^2z}{4\pi}\partial y \bar{\partial}y\,,\qquad y\sim y+2\pi r\,.
\end{align}
We conjecture that the dual worldsheet action is given by
\begin{align}\label{eq:JTbaraction}
\begin{aligned}
S_{J\bar{T}}=S_{\rm sym}-\frac{\mu}{\pi}\int \frac{d^2z}{2\pi}\bar{\beta}\partial y\,.
\end{aligned}
\end{align}

Unlike the $T\bar{T}$-deformation, the action is linear in $\beta$'s. Thus integrating out $\beta$'s gives a delta function that localizes the worldsheet moduli. As a result, we get
\begin{align}
\mathcal{Z}_{\rm WS}=\sum_{K>0}p^{K}\mathcal{T}_{K}[\mathcal{Z}_{+}]+p^{-K}\mathcal{T}_{K}[\mathcal{Z}_{-}]\,,
\end{align}
with
\begin{align}\label{eq:JTbarZpm}
\mathcal{Z}_{\pm}(R,\zeta,\bar{\zeta})=\pm \sum_{m,w} \frac{\mathcal{Z}_{\rm seed}^{(m,w)} \left(\pm \zeta, \frac{\pm\bar{\zeta}\,\mp\,\frac{\mu}{\pi}\frac{r}{R}m}{1\,\mp\,\frac{\mu}{\pi}\frac{r}{R}w}\right)} {1+\frac{\mu}{\pi}\frac{r}{R} \frac{m\,\mp w\,\zeta}{\zeta-\bar{\zeta}} } \, ,
\end{align}
where $Z^{(m,w)}_{\rm seed}$ is the seed CFT partition function in the $(m,w)$-winding sector of $y$. Expanding $\mathcal{Z}_{\rm seed}^{(m,w)}$ in $\mathcal{Z}_{\pm}$ \eqref{eq:JTbarZpm}, we obtain the spectrum,
\begin{align}\label{eq:JTbarpos}
&E_{R (+)}(\mu)=\frac{\pi^2}{\mu^2}\left[\left(R-\frac{\mu Q}{\pi}\right)-\sqrt{\left(R-\frac{\mu Q}{\pi}\right)^2-\frac{2\mu^2 R}{\pi^2}E_{R}}\right]\nonumber\\
&P_{(+)}(\mu)=P\,,
\end{align}
and 
\begin{align}\label{eq:JTbarneg}
&E_{R (-)}(\mu)=\frac{\pi^2}{\mu^2}\left[\left(R-\frac{\mu Q}{\pi}\right)+\sqrt{\left(R-\frac{\mu Q}{\pi}\right)^2-\frac{2\mu^2 R}{\pi^2}E_{R}}\right]\nonumber \\
&P_{(-)}(\mu)=P\,,
\end{align}
where $E_R=(E-P)/2$ and $Q=\frac{n}{r}+\frac{wr}{2}$ is the $U(1)$ charge. The positive-winding spectrum \eqref{eq:JTbarpos} agrees with the $J\bar{T}$-deformed spectrum obtained in the literature \cite{Frolov:2019xzi,Anous:2019osb,Chakraborty:2023wel} while the negative-winding spectrum \eqref{eq:JTbarneg} gives non-perturbative corrections.

\section{Conclusion}
We proposed string duals to 2d chiral Yang-Mills and to symmetric product orbifolds of general seed CFTs with $c<24$, including their $T\bar{T}$- and $J\bar{T}$-deformations. We presented supporting evidence, but further tests---such as higher-point amplitudes in 2d YM and correlation functions in symmetric orbifolds---are desirable \cite{toappear, toappear2}.

It is also important to clarify connections with other proposals for 2d YM duals, particularly the topological string dual to $q$-deformed YM \cite{Aganagic:2004js}. Such comparisons may also lead to improved formulations of our composite linear dilaton action.

A notable feature of our proposed duals for $T\bar{T}$- and $J\bar{T}$-deformations is the inclusion of negative winding sectors. These correspond to alternative sign choices in square roots and give non-perturbative corrections to the partition function. In nonrelativistic string, they correspond to modes that decouple in the nonrelativistic limit. As noted in \cite{Benjamin:2023nts}, such modes are crucial for the S-duality of the partition function. Understanding their role in $T\bar{T}$-deformations---possibly via resurgence analysis \cite{Gu:2024ogh,Griguolo:2022xcj,Griguolo:2022hek}---may be an interesting future direction.

Our proposal shares similarities with matrix string conjecture \cite{Motl:1997th, Dijkgraaf:1997vv}, which relates the discrete lightcone quantization (DLCQ) of non-perturbative type IIA superstring to 2d $\mathcal{N}=8$ super Yang-Mills (SYM). In this context, the perturbative IIA string corresponds to the IR limit of SYM, described by a symmetric product orbifold of supersymmetric $\mathbb{R}^{8}$. 
Meanwhile, the T-dual of the DLCQ string \cite{Harmark:2017rpg,Bergshoeff:2018yvt} is nonrelativistic string \cite{Gomis:2000bd,Danielsson:2000gi,Danielsson:2000mu} described by a $\beta$-$\gamma$ system. Together, these imply a duality between the symmetric orbifold of $\mathbb{R}^8$ and nonrelativistic type IIA string---a supersymmetric counterpart of our proposal. In addition, similarities between matrix string and 2d YM were discussed in interesting papers \cite{Billo:1998fb,Billo:1999bv}. Exploring these webs of dualities (cf.~\cite{Blair:2023noj}) further would be highly illuminating. 

\vspace{10pt}
\noindent 
\textbf{Acknowledgment} We thank 
Ashoke Sen for helpful suggestions. We also thank
Ofer Aharony, Alejandra Castro, Lorenz Eberhardt,  Matthias Gaberdiel,  Rajesh Gopakumar, Victor Gorbenko, David Gross, Bob Knighton, Jorrit Kruthorff, Suman Kundu, Juan Maldacena, Kiarash Naderi, Beat Nairz, Tal Sheaffer, Vit Sriprachyakul, Cumrun Vafa, Spenta Wadia and Ziqi Yan for discussions.

\bibliographystyle{apsrev4-1}
\bibliography{ref}

\clearpage

\onecolumngrid
\appendix

{\begin{center}\bf \Large{Supplemental Material}\end{center} }

\section{Chiral Composite Linear Dilaton: conformal invariance and comparison with the literature}
\subsubsection{Conformal invariance}
In this appendix, we show the conformal invariance of the chiral composite linear dilaton action \eqref{eq:CCLDaction}. We first write down the covariant form of the action in two parts:
\begin{equation}
    \begin{split}
       & S_{CLD}=S_{CLD,\,\text{kinetic}}+S_{CLD,\,\text{curvature}}\, ,\\
       & S_{CLD,\,\text{kinetic}} =\int\frac{d^2\sigma}{2\pi}\sqrt{-g}\, g^{ab}\,\partial_a\varphi\,\partial_b\varphi,\quad S_{CLD,\,\text{curvature}}=\int \frac{d^2\sigma}{\pi}\sqrt{-g}\, \hat{R}\,\varphi \, .
    \end{split}
\end{equation}
where 
\begin{equation}
    \varphi = \log\left[ \frac{1}{4} \, g^{ab}\partial_a X^{+}\partial_b X^{-}+\frac{i}{4}\,\epsilon^{cd} \,\partial_c X^{+}\partial_d X^{-} \right]\, .
\end{equation}
We have omitted an overall factor of $q/2$ which will be multiplied at the end. We note down variations of different quantities with respec to the metric $g^{ab}$ that will be useful for our computations.
\begin{itemize}
    \item Variation of metric determinant: \begin{equation}
        \frac{\delta\,(\sqrt{-g})}{\delta \,g^{ab}} =-\frac{1}{2}\sqrt{-g}\, g_{ab} \, .
    \end{equation}
    \item Variation of Ricci scalar: \begin{equation}
    \begin{split}
        & \delta \,\hat{R}=R_{ab}\,\delta g^{ab}+g^{ab}\,\delta R_{ab} =R_{ab}\,\delta g^{ab}+\nabla_c\,[g_{ab}\nabla^c(\delta g^{ab})]-\nabla_c\nabla_d\, (\delta g^{cd})\\
        \Rightarrow \, & \frac{\delta\,\hat{R}}{\delta g^{ab}}= \left( R_{ab}+\nabla_c\,g_{ab}\, \nabla^c-\nabla_a\nabla_b \right)=\left( \frac{1}{2}\hat{R}\, g_{ab}+g_{ab}\,\nabla_c\nabla^c-\nabla_a\nabla_b\right) \, . 
    \end{split}
\end{equation}

\item Variation of $\varphi$: Using $\frac{\delta \epsilon^{cd}}{\delta g^{ab}}=\frac{1}{2}\epsilon^{cd}g_{ab}$,
\begin{equation}
    \begin{split}
           & \frac{\delta\,\varphi}{\delta\, g^{ab}}=\frac{\frac{1}{4} \,\partial_a X^{+}\partial_b X^{-}+\frac{i}{4}\left( \frac{1}{2}\epsilon^{cd}g_{ab}\right)\, \partial_c X^{+}\partial_d X^{-}}{\frac{1}{4} \, g^{ab}\partial_a X^{+}\partial_b X^{-}+\frac{i}{4}\,\epsilon^{ab} \partial_a X^{+}\partial_b X^{-}} \\
        \Rightarrow\; & g^{ab} \frac{\delta\,\varphi}{\delta\, g^{ab}}=\frac{\frac{1}{4} \,g^{ab}\partial_a X^{+}\partial_b X^{-}+\frac{i}{4}\,\epsilon^{cd}\, \partial_c X^{+}\partial_d X^{-} \left( \frac{1}{2}g^{ab}g_{ab}\right)}{\frac{1}{4} \, g^{ab}\partial_a X^{+}\partial_b X^{-}+\frac{i}{4}\,\epsilon^{ab} \partial_a X^{+}\partial_b X^{-}} =1 \, .
    \end{split}
\end{equation}
\end{itemize}
The contribution to stress tensor from $S_{CLD,\,\text{kinetic}}$ is
\begin{equation}
    \begin{split}
        \left( T_{CLD,\,\text{kinetic}} \right)_{ab}&=-\frac{4\pi}{\sqrt{-g}}\frac{\delta S_{CLD,\,\text{kinetic}}}{\delta g^{ab}}\\
        &= -\frac{4\pi}{\sqrt{-g}} \left[ \frac{1}{2\pi}\left(-\frac{1}{2}\sqrt{-g}\, g_{ab}\right)\, g^{cd}\,\nabla_c \varphi\, \nabla_d\varphi+\frac{\sqrt{-g}}{2\pi}\, \nabla_a\varphi\,\nabla_b \varphi +2 \frac{\sqrt{-g}}{2\pi}\, g^{cd}\,\nabla_c\left( \frac{\delta\,\varphi}{\delta\, g^{ab}}\right)\nabla_d\varphi \right]\\
        &= -2\left[-\frac{1}{2}g_{ab}\,(\nabla\varphi)^2+ \nabla_a\varphi\,\nabla_b \varphi-2  \frac{\delta\,\varphi}{\delta\, g^{ab}}\, \nabla^2\varphi \right] \, .
    \end{split}
\end{equation}
where we have ignored the total derivative term which can be absorbed by a field redefinition. Now taking trace, i.e contracting $\left( T_{CLD,\,\text{kinetic}} \right)_{ab}$ with $g^{ab}$, we obtain
\begin{equation}
    \begin{split}
&  \left( T_{L,\,\text{kinetic}} \right)^a_{\;\;a}=-2\left[ -\frac{1}{2}g^{ab}g_{ab}\,(\nabla\varphi)^2+\,(\nabla\varphi)^2-2 \,g^{ab} \frac{\delta\,\varphi}{\delta\, g^{ab}} \, \nabla^2\varphi\right]=+4\,\nabla^2\varphi \, .
\end{split}
\end{equation}
The stress tensor contribution from $S_{CLD,\,\text{curvature}}$ is
\begin{equation}
    \begin{split}
        \left( T_{CLD,\,\text{curvature}}\right)_{ab}&=-\frac{4\pi}{\sqrt{-g}}\frac{\delta S_{CLD,\,\text{curvature}}}{\delta g^{ab}} \\
        & =-\frac{4\pi}{\sqrt{-g}} \left[ \frac{1}{\pi} \left(-\frac{1}{2}\sqrt{-g}\, g_{ab} \right)\, \hat{R}\,\varphi+\frac{\sqrt{-g}}{\pi}\frac{\delta \hat{R}}{\delta g^{ab}}\,\varphi+ \frac{\sqrt{-g}}{\pi}\,\hat{R}\,  \frac{\delta\,\varphi}{\delta\, g^{ab}} \right]\\
        &=-2\left[ -g_{ab}\,\hat{R}\,\varphi  +2\left( \frac{1}{2}\hat{R}\, g_{ab}+g_{ab}\,\nabla_c\nabla^c-\nabla_a\nabla_b\right) \varphi+ 2\,\hat{R}\,  \frac{\delta\,\varphi}{\delta\, g^{ab}}\right]\\
        &= 4\left[ \nabla_a\nabla_b\,\varphi -g_{ab}\,\nabla^2\varphi -\hat{R}\,  \frac{\delta\varphi}{\delta g^{ab}}  \right] \, .
    \end{split}
\end{equation}
Taking trace, it would contribute
\begin{equation}\label{eq:TCLDcurv}
    \begin{split}
          \left( T_{CLD,\,\text{curvature}}\right)^{a}_{\;\; a}&=4\left[ g^{ab}\,\nabla_a\nabla_b\,\varphi -(g^{ab}g_{ab})\,\nabla^2\varphi -g^{ab}\,\hat{R}\,  \frac{\delta\varphi}{\delta g^{ab}}  \right]\\
          &= -4\,\nabla^2\varphi-4\hat{R}\, .
    \end{split}
\end{equation}
Thus the trace of the stress tensor from the chiral composite dilaton action is
\begin{equation}
   \left( T_{CLD} \right)^a_{\;\; a}= \left( T_{CLD,\,\text{kinetic}} \right)^a_{\;\;a}+  \left( T_{CLD,\,\text{curvature}}\right)^{a}_{\;\; a}=-4\hat{R} \, .
\end{equation}
In the full worldsheet action, we have $\frac{q}{2} S_{CLD}$, and  so expressing $\frac{q}{2}\left( T_{CLD} \right)^a_{\;\; a}=-\frac{c_{\rm CLD}}{12}\hat{R}$, we can read off the central charge from the CLD part of our worldsheet
\begin{equation}
    c_{\rm CLD}=24q\,.
\end{equation}
which, together with $c=2$ coming from the $\beta$-$\gamma$ action, reproduces what we obtained from the stress tensor OPEs in \eqref{TT_OPE}. 

\subsubsection{Comparison with the literature}
The actions similar to $\mathcal{L}_{CLD}$ were proposed in the past:
\begin{itemize}
\item In \cite{Baba:2009ns}, a similar action with $\tilde{\varphi} = \partial X^{+}\bar{\partial} X^{+}$ (instead of $\varphi = \partial X^{+}\bar{\partial} X^{-}$) was studied in order to perform the dimensional regularization of string field theory in the lightcone gauge. (Precisely speaking, the curvature coupling $\hat{R}\tilde{\varphi}$ was not introduced in \cite{Baba:2009ns}, but it was implicitly used in the computation.) A counterpart in the operator formalism appeared in \cite{Kazama:2010ys}.
\item $\mathcal{L}_{\rm CLD}$ without the kinetic term was proposed in \cite{Callebaut:2019omt} as a worldsheet description for the $T\bar{T}$-deformed CFT.
\end{itemize}
Let us also make a brief comment on the proposal \cite{Callebaut:2019omt}. Their action lacks the kinetic term, leading to a non-traceless stress tensor due to the $-4\,\nabla^2\varphi$ term (cf.~\eqref{eq:TCLDcurv}). As a result, the renormalization group generates $\nabla^2\varphi$ term in the Lagrangian, indicating a violation of the conformal invariance. That said, the violation is mild since $\nabla^2 \varphi$ is proportional to the equation of motion, and can be removed by a field redefinition as is well-known in effective field theory. However, this redefinition makes fields to transform highly nonlinearly under conformal transformations, complicating the analysis---in particular in constructing conformally covariant vertex operators. We thus believe that some of the analyses in \cite{Callebaut:2019omt} cannot be fully justified.

\section{Torus partition function of worldsheet dual to chiral 2d YM}	
In this appendix, we provide technical details of the computation of the torus partition function fo string dual to chiral $2d$ YM.

We start with the action on the torus, 
\begin{equation}
S_{\mathds{T}^2}=\int \frac{d^2z}{2\pi} \left[\beta \, \bar{\partial}X+\bar{\beta}\,\partial\bar{X}+\frac{\lambda \pi }{2}\, (\partial X \bar{\partial}\bar{X}-\bar{\partial}X \partial \bar{X})+q\,\partial \varphi\, \bar{\partial}\varphi\right]+\text{bc ghost}\,.
\end{equation}
The partition function is given by
\begin{equation}
\mathcal{Z}_{\rm WS}=Z_{\beta, \bar{\beta}, X,\bar{X}}\,Z_{\text{bc}}\,,
\end{equation}
where 
\begin{equation}
Z_{\beta,\bar{\beta}, X,\bar{X}}=\int_{\mathcal{F}}\frac{d^2\tau}{\tau_2}\,\int [\mathcal{D}X\, \mathcal{D}\bar{X}\,\mathcal{D}\beta\,\mathcal{D}\bar{\beta}]\, e^{-S_{\beta,\bar{\beta},X,\bar{X}}} \,.
\end{equation}

\subsubsection{Cancellation between $bc$ ghosts and oscillator modes of $\beta$'s}
The contribution from $bc$ ghosts is given by a standard expression
\begin{equation}
    Z_{\text{bc}} = \int [\mathcal{D}b\,\mathcal{D}\bar{b}\, \mathcal{D}c\, \mathcal{D}\bar{c}] \, e^{-\int\frac{d^2z}{2\pi}\left(b\,\bar{\partial}c +\bar{b}\,\partial \bar{c}\right)}= \left[\text{det}_{\mathds{T}^2}\left( \frac{\bar{\partial}}{2\pi}\right) \right]\, \left[\text{det}_{\mathds{T}^2}\left( \frac{\partial}{2\pi}\right) \right]=|\eta(\tau)|^4\,.
\end{equation}

The path integral over matter sector splits into zero modes $\{X_0,\bar{X}_0,\beta_0,\bar{\beta}_0\}$ of $X,\bar{X}, \beta,\bar{\beta}$, and their oscillatory modes $\{ X_{osc},\bar{X}_{osc},\beta_{osc},\bar{\beta}_{osc} \} $ :  
\begin{equation}
\int [\mathcal{D}X\, \mathcal{D}\bar{X}\,\mathcal{D}\beta\,\mathcal{D}\bar{\beta}] \equiv \int dX_0 d\bar{X}_0\, [dX_{osc}][d\bar{X}_{osc}]\, \frac{d\beta_0 }{2\pi}\frac{d\bar{\beta}_0}{2\pi} \, [d\beta_{osc}] [d\bar{\beta}_{osc}] \, .
\end{equation}

To proceed, we first integrate out the oscillator modes of $\beta$'s. This gives
\begin{align}
\begin{aligned}
   & \int [d\beta_{osc}] [d\bar{\beta}_{osc}] \, \exp\left[ -\int \frac{d^2 z}{2\pi} \,\left( \beta_{osc} \,\bar{\partial} X_{osc} + \bar{\beta}_{osc}\, \partial \bar{X}_{osc}\right) \right] = \delta \left( \frac{\bar{\partial}X_{osc}}{2\pi}   \right) \, \delta \left( \frac{\partial\bar{X}_{osc}}{2\pi}  \right) \\
   & = \left[\text{det}_{\mathds{T}^2}\left( \frac{\bar{\partial}}{2\pi}\right) \right]^{-1}\, \left[\text{det}_{\mathds{T}^2}\left( \frac{\partial}{2\pi}\right) \right]^{-1} \, \delta\left(X_{osc}^{\text{anti-holo}} \right)\, \delta\left(\bar{X}_{osc}^{\text{holo}} \right)\,,
\end{aligned}   
\end{align}
where $\delta\left(X_{osc}^{\text{anti-holo}} \right)$ sets the oscillator part of $X$ to be purely holomorphic (and $\delta (\bar{X}_{osc}^{holo})$ is defined in a similar manner). Furthermore, the inverse determinants cancel the contributions from the $bc$ ghosts.

\subsubsection{Zero mode integrals and localization} Because of the delta functions above, $X$ ($\bar{X}$) needs to be (anti-)holomorphic except for zero modes. To be consistent with the periodicity of the target space and the worldsheet, $X$ and $\bar{X}$ need to take the following form:
\begin{equation}\label{eqap:2dYMzeromode}
    \begin{split}
        & \frac{1}{R}\,X(z,\bar{z}) = \frac{x}{R}+\frac{(m^1-\bar{\tau}\,w^1)+\zeta\, (m^2-\bar{\tau}\, w^2)}{\tau-\bar{\tau}}\,z+ \frac{(\tau \,w^1-m^1)+\zeta\,(\tau\, w^2-m^2)}{\tau-\bar{\tau}}\, \bar{z}\,,\\
        & \frac{1}{R}\, \bar{X}(z,\bar{z})= \frac{\bar{x}}{R}+\frac{(m^1-\bar{\tau}\,w^1)+\bar{\zeta}\, (m^2-\bar{\tau}\, w^2)}{\tau-\bar{\tau}}\,z+ \frac{(\tau \,w^1-m^1)+\bar{\zeta}\,(\tau\, w^2-m^2)}{\tau-\bar{\tau}}\, \bar{z}\,.
        \end{split}
\end{equation}
Here $x$ and $\bar{x}$ are constant modes taking values in the target space torus while $w^{1,2}$ and $m^{1,2}$ are integer winding numbers defined by
\begin{equation}\label{eqap:2dYMperiodicities}
\begin{split}
&X^1(z+2\pi,\bar{z}+2\pi)=X^1(z,\bar{z})+2\pi R \, w^1+2\pi R \, \zeta_1 w^2,\quad X^2(z+2\pi,\bar{z}+2\pi)=X^2(z,\bar{z})+2\pi R \, \zeta_2 w^2  \, , \\
& X^1(z+2\pi\tau,\bar{z}+2\pi\bar{\tau})=X^1(z,\bar{z})+2\pi R \, m^1+2\pi R \,\zeta_1 m^2, \quad  X^2(z+2\pi\tau,\bar{z}+2\pi \bar{\tau})=X^2(z,\bar{z})+2\pi R \,\zeta_2 m^2 \, .
\end{split}	
\end{equation}
Note that the delta functions alone do not forbid holomorphic and periodic terms for $X$ (and anti-holomorphic and periodic terms for $\bar{X}$). However there are no such functions (without singularities) on a torus, and this is why \eqref{eqap:2dYMzeromode} is justified.

Let us also note that one immediate consequence of \eqref{eqap:2dYMzeromode}: Since $X$ and $\bar{X}$ are linear in $z$ and $\bar{z}$, the kinetic term of $\mathcal{L}_{CLD}$ vanishes
\begin{align}
    \partial \varphi \bar{\partial} \varphi =\frac{\partial^2X}{\partial X} \frac{\bar{\partial}^2\bar{X}}{\bar{\partial} \bar{X}}=0\,.
\end{align}

We next perform the $\int d\beta_0 \, d\bar{\beta}_0$ integral. Using
\begin{equation}
	\bar{\partial} X = R \,\frac{(\tau w^1-m^1)+\zeta(\tau w^2-m^2)}{\tau-\bar{\tau}}\,.
\end{equation} 
we can evaluate the relevant part of the worldsheet action as
\begin{equation}
 \int \frac{d^2z}{2\pi} \,\beta_0 \, \bar{\partial} X =-2\pi i \beta_0 R\,(w^1+\zeta w^2)  \left(\tau-\frac{m^1+\zeta m^2}{w^1+\zeta w^2}\right)\,.
\end{equation} 
where we used $\int d^2z=2\int d^2\sigma=8\pi^2\tau_2$. Similarly from the part with $\bar{\beta}_0$, we obtain
\begin{equation}
	 \int \frac{d^2z}{2\pi} \,\bar{\beta}_0 \, \partial \bar{X} =2\pi i \bar{\beta_0}R\,(w^1+\bar{\zeta}w^2)\left(\bar{\tau}-\frac{m^1+\bar{\zeta}m^2}{w^1+\bar{\zeta}w^2}\right)\,.
\end{equation}
Performing the path-integral over $(\beta_0,\bar{\beta}_0)$ in the path integral, we will get a two-dimensional delta function:
\begin{equation}
\int \frac{d\beta_0}{2\pi}\,\frac{ d\bar{\beta}_0}{2\pi}\, [...] =	\frac{1}{4\pi^2 R^2\,|w^1+\zeta w^2|^2}\,\delta^{(2)}\left[\tau-\frac{m^1+\zeta m^2}{w^1+\zeta w^2}\right] \, ,
\end{equation}
which localizes the worldsheet modulus $\tau$ to be
\begin{equation}\label{localizing_points}
\tau_{*}=\frac{m^1+\zeta m^2}{w^1+\zeta w^2}\,.
\end{equation}

\subsubsection{Sum over winding numbers}
Thanks to the localization, the path integral can be fully performed and it reduces to an integral over the constant mode $x$ and $\bar{x}$, which gives
\begin{equation}
\int dx\, d\bar{x}=4\pi^2 R^2 \,\zeta_2\,.
\end{equation}
and a sum over winding numbers $w^{a}$ and $m^{a}$. Thus, putting everything together, we obtain
\begin{align}\label{eqap:2dYMintegral}
\mathcal{Z}_{\rm WS}=\!\!\sum_{\{m^{a},w^{a}\}}\frac{\zeta_2}{|W|^2}\int_{\mathcal{F}}\frac{d^2\tau}{\tau_2}\delta^{(2)}\left(\tau-\dfrac{M}{W}\right)e^{-S_{m,w}}\,,
\end{align}
where $M$ and $W$ are given by
\begin{align}
M=m^{1}+\zeta m^{2}\,,\qquad W=w^{1}+\zeta w^{2}\,,
\end{align}
and $S_{\{m^{a},w^{a}\}}$ is 
\begin{align}
        S_{m,w} = \left.\frac{\lambda\pi}{2}\int \frac{d^2z}{2\pi} \, \partial X\, \bar{\partial} \bar{X}\right|_{\text{on \eqref{eqap:2dYMzeromode}}}=\frac{\lambda \pi^2 R^2}{2\tau_2} \left|\bar{W}\tau-\bar{M}\right|^2 \, .
\end{align}
We will now exploit a trick \cite{Polchinski:1985zf} to simplify the sum over windings $\{m^a,w^a\}$. To do so, we first represent them in a matrix form
\begin{align}
\begin{pmatrix}
m^2 & m^1\\
w^2 & w^1
\end{pmatrix} \, .
\end{align}
We define $K$ to be the determinant of this matrix $K=(m^2w^1-w^2m^1)$. We then split the sum over windings in \eqref{worldsheet_partition_1} into sum over $K$ and matrices of the above form with fixed determinant $K$:
\begin{equation}
\mathcal{Z}_{\rm WS}(R,\zeta,\bar{\zeta})=\sum_{K\neq 0} \mathcal{Z}_K(R,\zeta,\bar{\zeta}) \, .
\end{equation}
As mentioned in the main text, the zero winding $K=0$ is excluded since the CLD action $\mathcal{L}_{CLD}$ diverges for $K=0$ and makes it vanish. 

To proceed, we use the fact that any matrix with integer entries and a fixed determinant can be obtained by applying an element of $SL(2,\mathbb{Z})$ to any of the matrices of the form
\begin{equation}\label{sl}
\mathcal{S}_{\pm K}=\left \{ \begin{pmatrix}
a&b\\ 
0& \pm d
\end{pmatrix},\quad a,b,d\in \mathds{Z}, \quad ad=K \in \mathbb{Z}_{+},\quad 0\leq b<d,\quad d>0\right\} \, .
\end{equation}
Here we choose the plus (minus) sign when the determinant is positive (negative).

\begin{figure}[!thb]
    \centering
\begin{tikzpicture} [scale = 1]



\draw[white, fill = cyan! 10! ] (-1,0) -- (1,0) -- (1,5) -- (-1,5) -- (-1,0);

\draw[white, fill = yellow! 15! ] (-3,0) -- (-1,0) -- (-1,5) -- (-3,5) -- (-3,0);

\draw[white, fill = brown! 10! ] (3,0) -- (1,0) -- (1,5) -- (3,5) -- (3,0);

\draw[white, fill = violet! 10! ] (-5,0) -- (-3,0) -- (-3,5) -- (-5,5) -- (-5,0);

\draw[white, fill = teal! 10! ] (5,0) -- (3,0) -- (3,5) -- (5,5) -- (5,0);



\draw[white, fill= orange! 15! ] (-2,0) arc (180:0:2) -- cycle;


\draw[white, fill= gray! 15! ] (-6,0) arc (180:0:2) -- cycle;


\draw[white, fill= lime! 15! ] (2,0) arc (180:0:2) -- cycle;


\draw[white, fill= white ] (-4,0) arc (180:0:2) -- cycle ;


\draw[white, fill= white ] (0,0) arc (180:0:2) -- cycle ;


\draw[thick] (-2,0) arc (180:0:2) -- cycle;


\draw[thick] (-6,0) arc (180:0:2) -- cycle;


\draw[thick] (2,0) arc (180:0:2) -- cycle;


\draw[thick] (-4,0) arc (180:0:2) -- cycle;


\draw[thick] (0,0) arc (180:0:2) -- cycle;


\draw[ thick] (0.657,0) circle (0.657);

\draw[ thick] (-0.657,0) circle (0.657);

\draw[ thick] (1.343,0) circle (0.657);

\draw[ thick] (-1.343,0) circle (0.657);

\draw[ thick] (0.657,0)+(2,0) circle (0.657);

\draw[ thick] (-0.657,0)+(-2,0) circle (0.657); 

\draw[ thick] (1.343,0)+(2,0) circle (0.657);

\draw[thick] (-1.343,0)+(-2,0) circle (0.657);

\draw[white, fill=white! 15] (-6,0) rectangle (6,-0.67);

\draw[white, fill=white! 15!] (-4,0) rectangle (-7,5);

\draw[white, fill=white! 15!] (4,0) rectangle (7,5);

\draw[thick] (-1,0) -- (-1,5) (1,0) -- (1,5) (3,0) -- (3,5) (-3,0) -- (-3,5);

\draw[thick, dashed] (-4,0) -- (4,0);



\node at (0,-0.5) {0};
\node at (2,-0.5) {1};
\node at (-2,-0.5) {-1};
\node at (4,-0.5) {2};
\node at (-4,-0.5) {-2};

\node at (0,3.5) {$\mathcal{F}$};
\node at (2,3.5) {$T$};
\node at (-2,3.5) {$T^{-1}$};
\node at (3.6,3.5) {$T^2$};
\node at (-3.6,3.5) {$T^{-2}$};

\node at (0,1.5) {\scalebox{0.8}{$S$}};
\node at (2,1.5) {\scalebox{0.8}{$TS$}};
\node at (-2,1.5) {\scalebox{0.8}{$T^{-1}S$}};
\node at (3.9,1.5) {\scalebox{0.8}{$T^2S$}};
\node at (-3.9,1.5) {\scalebox{0.8}{$T^{-2}S$}};

\node at (0.67,1) {\scalebox{0.6}{$ST^{-1}$}};
\node at (-0.67,1) {\scalebox{0.6}{$ST$}};

\node at (1.35,1) {\scalebox{0.6}{$TST$}};
\node at (-1.35,1) {\scalebox{0.6}{$STS$}};

\node at (2.65,1) {\scalebox{0.5}{$TST^{-1}$}};
\node at (-2.65,1) {\scalebox{0.5}{$T^{-1}ST$}};

\node at (3.4,1) {\scalebox{0.5}{$T^2ST$}};
\node at (-3.35,1) {\scalebox{0.4}{$T^{-1}STS$}};

\end{tikzpicture}
\caption{Extending the fundamental domain of the torus moduli using $SL(2,\mathbb{Z})$}
\end{figure}
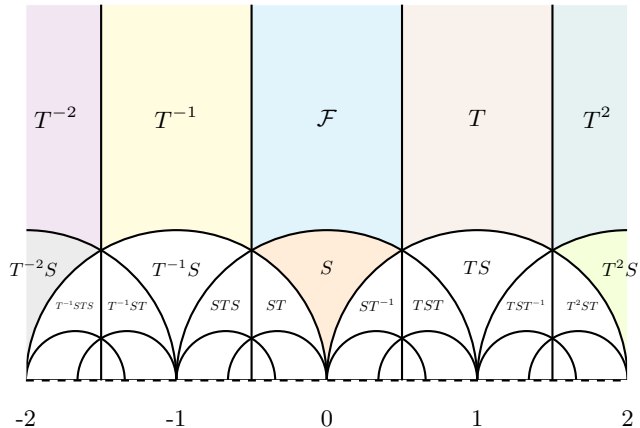

Using the above $SL(2,\mathds{Z})$, we can ``unfold" the fundamental domain of the $\tau$ integral \eqref{eqap:2dYMintegral} to the whole upper half plane $\mathcal{H}_{+}$ and replace the sum over $\{m^a,w^a\}$  as a sum over matrices of the form  (\ref{sl}) with fixed $(\pm K)$, i.e 
\begin{equation}\label{split}
	\mathcal{Z}(R,\zeta,\bar{\zeta})=\sum_{K=1}^{\infty}\mathcal{Z}_{\pm K}(R,\zeta,\bar{\zeta}) \, ,
\end{equation}
with 
\begin{equation}\label{1}
\mathcal{Z}_{\pm K}(R,\zeta,\bar{\zeta})=\zeta_2\sum_{\mathcal{S}_{\pm K}}\frac{1}{d^2}\int_{\mathcal{\mathcal{H}_{+}}} \frac{d^2\tau}{\tau_2} \,e^{-S^{\pm}_{\{a,b,d\}}}\,\delta^{(2)}\left(\tau\mp\frac{b+a\zeta}{d}\right)\,,
\end{equation}
and
\begin{equation}
S^{\pm}_{\{a,b,d\}}=\frac{1}{2}\lambda \frac{\pi^2R^2d^2}{\tau_2}\left|\tau\mp\frac{b+a\bar{\zeta}}{d}\right|^2\,.
\end{equation}

We quickly make an observation that, since $a,b,d>0$ in (\ref{sl}) and $\zeta$ being the target space  modulus satisfies $\zeta_2>0$, $(\tau_{*})_{-}=-\frac{b+a\zeta}{d}$ lies outside the integration region $\mathcal{H}_{+}$. Therefore, the integral for $\mathcal{Z}_{-K}$ vanishes:
\begin{equation}
\mathcal{Z}_{-K}(R,\zeta,\bar{\zeta})=0 \, .
\end{equation}  
On the other hand, the localization point $(\tau_*)_{+}=\frac{b+a\zeta}{d}$ lies inside $\mathcal{H}_{+}$ for $\mathcal{Z}_{+K}$. Evaluating $S^{+}_{\{a,b,d\}}$ at $(\tau_{\ast})_{+}$, we obtain
\begin{equation}
	\left.S^{+}_{\{a,b,d\}}\right|_{\tau=(\tau_*)_{+}}=\frac{1}{2}\lambda \, (4\pi^2R^2\zeta_2)\, ad= \frac{\lambda K  \text{Area}}{2}\,.
\end{equation} 
We therefore have
\begin{equation}
\begin{split}
\mathcal{Z}_{+ K}(R,\zeta,\bar{\zeta})=\frac{1}{K}\sum_{S_{+K}} e^{-\frac{\lambda K \text{Area}}{2} }= \frac{e^{-\frac{\lambda K \text{Area}}{2}}}{K}\sum_{ad=K}\sum_{b=0}^{d-1}1=\frac{e^{-\frac{\lambda K \text{Area}}{2}}}{K}\sum_{ad=K} d \, ,
\end{split}
\end{equation}
and so we conclude from (\ref{split}),
	\begin{equation}\label{worldsheet_one-loop}
	\mathcal{Z}_{\rm WS} =\sum_{K=1}^{\infty} \frac{e^{-\frac{\lambda K \text{Area}}{2}}}{2K}\sum_{ad=K} (a+d) \, ,
	\end{equation}	
where we used 
$$\sum_{ad=K} (a+d)=2\sum_{ad=K} d \, .$$
The result \eqref{worldsheet_one-loop} matches precisely with the leading expression of free energy in the chiral 2d YM.

\section{Torus partition function for $T\bar{T}$-deformation}
In this appendix, we explain the details of the computation of the torus partition function for the string dual to $T\bar{T}$-deformation.

As mentioned in the main text, the first step of the computation is to integrate out $\beta$'s. We then obtain the action
\begin{align}\label{eqap:TTbaraction}
 S=\int \frac{d^2z}{2\mu} \,\bar{\partial}X\, \partial \bar{X}+\frac{q}{2}\int\frac{d^2z}{2\pi}\mathcal{L}_{CLD} -B\int \frac{d^2 z}{4\pi} \left(\partial X \bar{\partial} \bar{X}-\bar{\partial}X\partial \bar{X}\right) +S_{\rm seed} +\text{bc ghost}\,.
\end{align}
Unlike the case with 2d YM and symmetric product orbifolds, the path integral does not localize to covering maps. This makes it difficult to perform the path integral in the conformal gauge due to the nonlinearity in $\mathcal{L}_{\rm CLD}$. Instead, we impose the winding gauge $\partial X= \bar{\partial}\bar{X}={\rm const}$. Imposing such a gauge fixing is common in the presence of nonlinearity in the action; cf. Green-Schwarz superstring and string in plane-wave backgrounds. An immediate consequence of the gauge-fixing is that the composite linear dilaton term $\mathcal{L}_{\rm CLD}$ does not contribute since $\partial\varphi \bar{\partial}\varphi=0$ for $\partial X=\bar{\partial}\bar{X}={\rm const}$.

After imposing the gauge condition, we obtain
\begin{align}\label{eqap:aftergauge}
\begin{aligned}
&X=\# z + \left[x+\#\bar{z}+(\text{anti-holomorphic, periodic})\right]\,,\\
&\bar{X}=\# \bar{z} + \left[\bar{x}+\#z+(\text{holomorphic, periodic})\right]\,,
\end{aligned}
\end{align}
Here the first terms on the right hand side come directly from the gauge-fixing conditions while the terms in the square brackets are the ones that are not killed by the gauge fixing. Since there are no holomorphic functions that are periodic and non-singular on a torus, we can further set the last terms in \eqref{eqap:aftergauge} to zero. We thus conclude that the winding gauge forces $X$ and $\bar{X}$ to be linear functions of $z$ and $\bar{z}$ up to constant modes $x$ and $\bar{x}$. 

Therefore, by imposing the winding gauge and rescaling the worldsheet coordinates to keep the standard periodicity $z\sim z+2\pi \sim z+2\pi \tau$, we obtain the following expressions for $X$ and $\bar{X}$ for each winding sector,
\begin{equation}\label{eqap:TTbarzeromode}
    \begin{split}
        & \frac{1}{R}\,X(z,\bar{z}) = \frac{x}{R}+\frac{(m^1-\bar{\tau}\,w^1)+\zeta\, (m^2-\bar{\tau}\, w^2)}{\tau-\bar{\tau}}\,z+ \frac{(\tau \,w^1-m^1)+\zeta\,(\tau\, w^2-m^2)}{\tau-\bar{\tau}}\, \bar{z}\,,\\
        & \frac{1}{R}\, \bar{X}(z,\bar{z})= \frac{\bar{x}}{R}+\frac{(m^1-\bar{\tau}\,w^1)+\bar{\zeta}\, (m^2-\bar{\tau}\, w^2)}{\tau-\bar{\tau}}\,z+ \frac{(\tau \,w^1-m^1)+\bar{\zeta}\,(\tau\, w^2-m^2)}{\tau-\bar{\tau}}\, \bar{z}\,.
        \end{split}
\end{equation}
Here $w^{a}$ and $m^{a}$ are winding numbers defined by \eqref{eqap:2dYMperiodicities}. 

We then compute the path integral by evaluating the action \eqref{eqap:TTbaraction} on the map \eqref{eqap:TTbarzeromode}. This leads to
\begin{align}\label{worldsheet_partition_TT-bar_1}
\begin{aligned}
\mathcal{Z}_{\rm WS}&=4\pi^2 R^2\, \zeta_2\,\sum_{\{m^a,w^a\}}p^{(m^2w^1-m^1w^2)}\int_{\mathcal{F}} \frac{d^2\tau}{\tau_2} \,e^{-S_{m,w}} \, \mathcal{Z}_{\rm seed}(\tau,\bar{\tau})\, (4\pi\mu\tau_2)^{-1}\\
&=\frac{\pi R^2\zeta_2}{\mu} \sum_{\{m^a,w^a\}} p^{(m^2w^1-m^1w^2)}\int_{\mathcal{F}} \frac{d^2\tau}{\tau_2^2} \,e^{-S_{m,w}} \, \mathcal{Z}_{\rm seed}(\tau,\bar{\tau}) \, ,
\end{aligned}
\end{align}
where 
\begin{align}
    \begin{aligned}
        S_{m,w} &= \left.\int \frac{d^2z}{2\mu}  \bar{\partial}X \partial \bar{X} \right|_{\text{on \eqref{eqap:TTbarzeromode}}}
        = \frac{\pi^2 R^2}{\mu\tau_2} \left|(m^1-\tau w^1)+\zeta\,(m^2-\tau w^2) \right|^2\,,\\
        p&=e^{B \,\text{Area}/(2\pi)}\,.
    \end{aligned}
\end{align}
Here the overall prefactor $4\pi^2 R^2 \zeta_2$ in \eqref{worldsheet_partition_TT-bar_1} comes from the integral over constant modes $x$ and $\bar{x}$ while the factor $(4\pi \mu\tau_2)^{-1}$ is a standard factor that arises in the one-loop string partition function and comes from a careful treatment of normalization of the path integral.

To proceed, we use the trick of extending the integration region to the upper-half plane $\mathcal{H}_{+}$ as in the analysis of 2d YM described above.
This leads to an expression
\begin{equation}
   \mathcal{Z}_{\rm WS} =\sum_{K>0} p^K\,\mathcal{Z}_K(R,\zeta,\bar{\zeta})+ p^{-K}\,\mathcal{Z}_{-K}(R,\zeta,\bar{\zeta})\,,
\end{equation}
where $\mathcal{Z}_{\pm K}$ is given by
\begin{equation}
\mathcal{Z}_{\pm K}( R,\zeta,\bar{\zeta})=\frac{\pi R^2\zeta_2}{\mu} \sum_{\substack{ad=K, \,d>0\\0\leq b<d}}\int_{\mathcal{\mathcal{H}_{+}}} \frac{d^2\tau}{\tau_2^2} \,e^{-\frac{\pi^2 d^2R^2}{\mu\tau_2} \left|\frac{a\zeta+b}{d}-\tau \right|^2}\, \mathcal{Z}_{\rm seed}(\tau,\bar{\tau})\,.
\end{equation}
Alternatively, we can express them using the Hecke operator as
\begin{align}
\mathcal{Z}_{\pm K}=\mathcal{T}_{K}\left[\mathcal{Z}_{\pm} (R,\zeta,\bar{\zeta})\right]\,,
\end{align}
where $\mathcal{T}_{K}$ is given by \eqref{eq:Hecke} and $\mathcal{Z}_{\pm}$ are $\mathcal{Z}_{K=\pm 1}$:
\begin{align}\label{eqap:Zpm}
\mathcal{Z}_{\pm}( R,\zeta,\bar{\zeta})=\frac{\pi R^2\zeta_2}{\mu} \int_{\mathcal{\mathcal{H}_{+}}} \frac{d^2\tau}{\tau_2^2} \,e^{-\frac{\pi^2 d^2R^2}{\mu\tau_2} \left|\zeta\mp \tau \right|^2}\, \mathcal{Z}_{\rm seed}(\tau,\bar{\tau})\,.
\end{align}

Plugging in the expansion of $\mathcal{Z}_{\rm seed}$ in terms of the undeformed spectrum
\begin{equation}
    \mathcal{Z}_{\rm seed}(\tau,\bar{\tau}) = \sum_n \exp[-2\pi \tau_2 RE_n+2\pi i\tau_1 RP_n]\,,
\end{equation}
to \eqref{eqap:Zpm}, we obtain the deformed spectra
\begin{align}
\begin{aligned}
&E_{n,(+)}(\mu)=\frac{\pi R}{\mu}\left(-1+\sqrt{1+\frac{\mu^2P_n^2}{\pi^2R^2}+\frac{2 \mu E_n}{\pi R}}\right)\,,\\ &P_{n,(+)}(\mu)=P_n\,,
\end{aligned}
\end{align}
and
\begin{align}
\begin{aligned}
&E_{n,(-)}(\mu)=\frac{\pi R}{\mu}\left(1+\sqrt{1+\frac{\mu^2P_n^2}{\pi^2 R^2}+\frac{2 \mu E_n}{\pi R}}\right)\,,\\ &P_{n,(-)}(\mu)=P_n\,.
\end{aligned}
\end{align}

\section{Torus partition function for $J\bar{T}$ deformation}
In this appendix, we provide computational details of the torus partition function dual to $J\bar{T}$ deformation.

The relevant worldsheet action is
\begin{equation}
S=\int \frac{d^2z}{2\pi} \,\Big[  \beta\, \bar{\partial}X+\bar{\beta}\,\partial \bar{X}-\frac{\mu}{\pi}\bar{\beta}\partial y\,+\frac{q}{2}\mathcal{L}_{CLD} -\frac{B}{2} \left(\partial X \bar{\partial} \bar{X}-\bar{\partial}X\partial \bar{X}\right)\Big]+S_{\rm seed} +\text{bc ghost} \, .
\end{equation}
The worldsheet partition function is given by
\begin{equation}
\mathcal{Z}=Z_{\beta, \bar{\beta}, X,\bar{X},y}\,Z_{{\rm CFT}^{\prime}}\,Z_{\text{bc}} \, ,
\end{equation}
where $Z_{{\rm CFT}^{\prime}}$ is a contribution from $S_{\rm seed}$ except for $y$ and
\begin{equation}
Z_{\beta,\bar{\beta}, X,\bar{X},y}=\int_{\mathcal{F}}\frac{d^2\tau}{\tau_2}\,\int [\mathcal{D}X\, \mathcal{D}\bar{X}\,\mathcal{D}y\,\mathcal{D}\beta\,\mathcal{D}\bar{\beta}]\, e^{-S_{\beta,\bar{\beta},X,\bar{X},Y}}  \, ,
\end{equation}
and 
\begin{equation}
    Z_{\text{bc}} = \int [\mathcal{D}b\,\mathcal{D}\bar{b}\, \mathcal{D}c\, \mathcal{D}\bar{c}] \, e^{-\int\frac{d^2z}{2\pi}\left(b\,\bar{\partial}c +\bar{b}\,\partial \bar{c}\right)}= \left[\text{det}_{\mathds{T}^2}\left( \frac{\bar{\partial}}{2\pi}\right) \right]\, \left[\text{det}_{\mathds{T}^2}\left( \frac{\partial}{2\pi}\right) \right]=|\eta(\tau)|^4 \, .
\end{equation}
The path integral over matter sector splits into zero modes $\{X_0,\bar{X}_0,\beta_0,\bar{\beta}_0,y_0\}$ of $X,\bar{X}, \beta,\bar{\beta},y$, and their oscillatory modes $\{ X_{osc},\bar{X}_{osc},\beta_{osc},\bar{\beta}_{osc},y_{osc} \} $ :  
\begin{equation}
\int [\mathcal{D}X\, \mathcal{D}\bar{X}\,\mathcal{D}y\,\mathcal{D}\beta\,\mathcal{D}\bar{\beta}] \equiv \int  dX_0 d\bar{X}_0 dy_0\, [dX_{osc}][d\bar{X}_{osc}] [dy_{osc}]\, \frac{d\beta_0 }{2\pi}\frac{d\bar{\beta}_0}{2\pi} \, [d\beta_{osc}] [d\bar{\beta}_{osc}]\,.
\end{equation}

\subsubsection{Performing the path integral}
To perform the path integral, we first decompose the path integral of $y$ into a given windng sector:
\begin{align}
y(z+2\pi,\bar{z}+2\pi)=y(z,\bar{z})+2 \pi r w\,,\qquad y(z+2\pi\tau,\bar{z}+2\pi\bar{\tau})=y(z,\bar{z})+2 \pi r m\,.
\end{align}
Next, for each winding sector, we integrate out $\beta_{osc}$'s:
\begin{align}
\begin{aligned}
   & \int [d\beta_{osc}] [d\bar{\beta}_{osc}] \, \exp\left[ -\int \frac{d^2 z}{2\pi} \,\left( \beta_{osc} \,\bar{\partial} X_{osc} + \bar{\beta}_{osc}\, (\partial \bar{X}_{osc}-\frac{\mu}{\pi}\partial y_{osc} )\right) \right] = \delta \left( \frac{\bar{\partial}X_{osc}}{2\pi}   \right) \, \delta \left( \frac{\partial (\bar{X}_{osc} -\frac{\mu}{\pi} y_{osc})}{2\pi}  \right) \\
   & = \left[\text{det}_{\mathds{T}^2}\left( \frac{\bar{\partial}}{2\pi}\right) \right]^{-1}\, \left[\text{det}_{\mathds{T}^2}\left( \frac{\partial}{2\pi}\right) \right]^{-1} \, \delta\left(X_{osc}^{\text{anti-holo}}\right)\, \delta\left((\bar{X}_{osc} -\frac{\mu}{\pi} y_{osc})^{\rm holo}\right)\,,
\end{aligned}   
\end{align}
where $\delta (f^{\text{ (anti-)holo}})$ sets the (anti-)holomorphic part of the field $f$ to zero. As bfore, the determinants in the prefactor cancel the contributions from $bc$ ghosts.

Owing to these delta functions, we find that $X$ and $\bar{X}-\frac{\mu}{\pi}y$ take the following form
\begin{align}
\begin{aligned}
 & \frac{1}{R}\,X(z,\bar{z}) = {\rm const.}+\frac{(m^1-\bar{\tau}\,w^1)+\zeta\, (m^2-\bar{\tau}\, w^2)}{\tau-\bar{\tau}}\,z+ \frac{(\tau \,w^1-m^1)+\zeta\,(\tau\, w^2-m^2)}{\tau-\bar{\tau}}\, \bar{z}\,,\\
        & \frac{1}{R}\, \left(\bar{X}(z,\bar{z})-\frac{\mu}{\pi}y\right)= {\rm const.}+\frac{(\tilde{m}^1-\bar{\tau}\,\tilde{w}^1)+\bar{\zeta}\, (m^2-\bar{\tau}\, w^2)}{\tau-\bar{\tau}}\,z+ \frac{(\tau \,\tilde{w}^1-\tilde{m}^1)+\bar{\zeta}\,(\tau\, w^2-m^2)}{\tau-\bar{\tau}}\, \bar{z}\,,
\end{aligned}
\end{align}
with $\tilde{m}^{1}=m^{1}-\frac{\mu r}{\pi R}m$ and $\tilde{w}^{1}=w^{1}-\frac{\mu r}{\pi R}w$.
Here we again used the fact that there are no holomorphic periodic functions on a torus that are non-singular. As in the case with 2d YM, this makes the contribution from $\mathcal{L}_{CLD}$ vanish since $\partial \varphi =0$.

Now, using these expressions, we next integrate out $\beta_0$'s. Using
\begin{equation}
 \int \frac{d^2z}{2\pi} \,\beta_0 \, \bar{\partial} X =-2\pi i \beta_0 R\,(w^1+\zeta w^2)  \left[\tau-\frac{m^1+\zeta m^2}{w^1+\zeta w^2}\right]\,,
\end{equation} 
and
\begin{equation}
    \int \frac{d^2z}{2\pi} \,\bar{\beta}_0 \, \left( \partial \bar{X}-\frac{\mu}{\pi}\partial y \right) = 2\pi i R\,\bar{\beta}_0\, \left(w^1+\bar{\zeta}\,w^2-\frac{\mu}{\pi}\frac{r}{R} w\right)\, \left[\bar{\tau}-\frac{m^1+\bar{\zeta}\, m^2-\frac{\mu}{\pi}\frac{r}{R} m}{w^1+\bar{\zeta} \,w^2-\frac{\mu}{\pi}\frac{r}{R} w}\right]\,,
\end{equation}
we obtain the delta functions for the worldsheet moduli
\begin{align}
\begin{aligned}
    &\int \frac{d\beta_0}{2\pi} \,e^{-\int \frac{d^2z}{2\pi} \,\beta_0 \, \bar{\partial} X} =\frac{1}{2\pi R\,(w^1+\zeta w^2)}\, \delta \left[\tau-\frac{m^1+\zeta m^2}{w^1+\zeta w^2}\right]\,,\\
     &\int \frac{d\bar{\beta}_0}{2\pi} \, e^{-  \int \frac{d^2z}{2\pi} \,\bar{\beta}_0 \, \left( \partial \bar{X}-\frac{\mu}{\pi}\partial y \right)} =\frac{1}{2\pi R\,\left(w^1+\bar{\zeta}\,w^2-\frac{\mu}{\pi}\frac{r}{R}w \right)}\, \delta\left[\bar{\tau}-\frac{m^1+\bar{\zeta}\, m^2-\frac{\mu}{\pi}\frac{r}{R} m}{w^1+\bar{\zeta} \,w^2-\frac{\mu}{\pi}\frac{r}{R} w}\right]\,.
     \end{aligned}
\end{align}
We are thus left with the following expression of the worldsheet partition function : 
\begin{equation}\label{worldsheet_partition_1}
\begin{split}
\mathcal{Z}_{\rm Ws}&= 4\pi^2 R^2\, \zeta_2\,\sum_{m,w}\sum_{\{m^a,w^a\}} p^{(m^2w^1-m^1w^2)}\int_{\mathcal{F}} \frac{d^2\tau}{\tau_2}   \frac{1}{4\pi^2 R^2 \, (w^1+\zeta w^2) \left(w^1+\bar{\zeta}\,w^2-\frac{\mu}{\pi}\frac{r}{R}w \right) } \, \delta \left[\tau-\frac{m^1+\zeta m^2}{w^1+\zeta w^2}\right]\\& \quad \quad \quad \quad  \quad \quad  \quad \quad  \quad \quad \times\; \delta\left[\bar{\tau}-\frac{m^1+\bar{\zeta}\, m^2-\frac{\mu}{\pi}\frac{r}{R} m}{w^1+\bar{\zeta} \,w^2-\frac{\mu}{\pi}\frac{r}{R} w}\right]\, \mathcal{Z}_{\rm seed}^{(m,w)} (\tau,\bar{\tau}) \,,
\end{split}
\end{equation}
where $\mathcal{Z}^{(m,w)}_{\rm seed}$ is the seed CFT partition function in the $(m,w)$ winding sector: 
\begin{align}
\mathcal{Z}_{\rm seed}^{(m,w)}=\mathcal{Z}_{\mathbb{S}^{1}}^{(m,w)}\times \mathcal{Z}_{{\rm seed}^{\prime}}\,,
\end{align}
where $Z_{{\rm seed}^{\prime}}$ is the partition function of the seed CFT without $y$ and $Z_{\mathbb{S}^{1}}$ is
\begin{equation}
\begin{split}
    \mathcal{Z}_{\mathbb{S}^1}^{(m,w)}(\tau,\bar{\tau}) &= \frac{\int_0^{2\pi r} dy_{const}}{(4\pi^2\cdot 2\tau_2)^{1/2}|\eta(\tau)|^2} \exp\left[- \frac{\pi r^2}{2\tau_2}|m-\tau \,w|^2 \right]\\
    &= \frac{r}{\sqrt{2\tau_2} \,|\eta(\tau)|^2}\,\exp\left[- \frac{\pi r^2}{2\tau_2}|m-\tau \,w|^2 \right]\,.
\end{split}    
\end{equation}

\subsubsection{Sum over windings}
We then use the same trick as before; extending the integration region to the upper half-plane using a sum over windings $(w^{a}, m^{a})$. After that, we compute the moduli integral $d^2\tau$ explicitly using the delta functions. Strictly speaking, this step is subtle since the delta functions do not have support on the upper half plane $\mathcal{H}_{+}$ because of the shifts $-\frac{\mu r}{\pi R} (m,w)$. A more proper way to perform the computation is to introduce a small $T\bar{T}$ coupling, make the integral of $\tau$ Gaussian, compute the integral and set the $T\bar{T}$ coupling to zero in the end. In practice, this amounts to analytically continuing the integration contours of $\tau$. 

After these technical manipulations, we find that the partition function receives contributions from both positive and negative winding sectors:
\begin{equation}
   \mathcal{Z}_{\rm WS} =\sum_{K>0} p^K\,\mathcal{Z}_K(R,\zeta,\bar{\zeta})+ p^{-K}\,\mathcal{Z}_{-K}(R,\zeta,\bar{\zeta}) \, ,
\end{equation}
where
\begin{equation}
    \mathcal{Z}_{\pm K}(R,\zeta,\bar{\zeta}) = \zeta_2\sum_{m,w}  \sum_{\substack{ad =K\\0\leq b<d, \,d>0}} \frac{1}{(\tau_2)_*} \frac{1}{d\left( d\mp \frac{\mu}{\pi}\frac{r}{R}\,w\right)}\, \mathcal{Z}_{\rm seed}^{(m,w)} (\tau_*,\bar{\tau}_*)  \, ,
\end{equation}
with
\begin{equation}
    \tau_*= \pm \frac{b+a\,\zeta}{d},\quad \bar{\tau}_*= \frac{b+a\, \bar{\zeta}-\frac{\mu}{\pi}\frac{r}{R}m}{\pm d-\frac{\mu}{\pi}\frac{r}{R}w},\; \Rightarrow \;(\tau_2)_*=\frac{a\,\zeta_2-\frac{i}{2}\frac{\mu}{\pi} \frac{r}{R} \left(m\mp w \frac{a\,\zeta+b}{d} \right) }{ \pm d-\frac{\mu}{\pi}\frac{r}{R}w}\,.
\end{equation}
With this, we can verify that it admits an expression in terms of the Hecke operator

\begin{equation}
\mathcal{Z}_{\pm K}(R,\zeta,\bar{\zeta})=\mathcal{T}_{K>0} \left[ \mathcal{Z}_{\pm }(R,\zeta,\bar{\zeta})\right] \, ,
\end{equation}
with 
\begin{align}
\mathcal{Z}_{\pm}(R,\zeta,\bar{\zeta})=\pm \sum_{m,w} \frac{\mathcal{Z}_{\rm seed}^{(m,w)} \left(\pm \zeta, \frac{\pm\bar{\zeta}\,\mp\,\frac{\mu}{\pi}\frac{r}{R}m}{1\,\mp\,\frac{\mu}{\pi}\frac{r}{R}w}\right)} {1+\frac{\mu}{\pi}\frac{r}{R} \frac{m\,\mp w\,\zeta}{\zeta-\bar{\zeta}} }\,.
\end{align}

\subsubsection{Reading off the $J\bar{T}$-deformed spectrum}To read off the spectrum, we expand $Z_{{\rm seed}^{\prime}}$ as 
\begin{equation}
    \begin{split}
        \mathcal{Z}_{{\rm seed}^{\prime}}(\tau,\bar{\tau}) &= \sum_s \exp[-2\pi\tau_2\cdot RE_s+i\cdot 2\pi \tau_1\cdot R P_s]= \sum_s \exp[ -4\pi\tau_2\cdot R\,(E'_R)_s +i\cdot 2\pi\tau \cdot P'_s  ] \, ,
    \end{split}
\end{equation}
with $E_R=(E-P)/2$. We next absorb the oscillator contribution of $\mathcal{Z}_{\mathbb{S}^{1}}$ into $ \mathcal{Z}_{{\rm seed}^{\prime}}(\tau,\bar{\tau})$ and make a new label $k$ of quantum numbers that encode both the states of ${\rm seed}^{\prime}$ and $\mathbb{S}^1$-excitations:
\begin{equation}
    \frac{1}{|\eta(\tau)|^2} \mathcal{Z}_{{\rm seed}^{\prime}}(\tau,\bar{\tau}) = \sum_k \exp[ -4\pi\tau_2\cdot R\,(E'_R)_k +i\cdot 2\pi\tau \cdot P'_k  ]\,.
\end{equation}
We then have
\begin{equation}
\begin{split}
    \mathcal{Z}^{(m,w)}_{{\rm seed}}(\tau,\bar{\tau}) &= \frac{r}{\sqrt{2\tau_2} \,|\eta(\tau)|^2}\,\exp\left[- \frac{\pi r^2}{2\tau_2}|m-\tau \,w|^2 \right]\, \mathcal{Z}_{{\rm seed}^{\prime}}(\tau,\bar{\tau})  \\&=\frac{r}{\sqrt{2\tau_2}} \exp\left[- \frac{\pi r^2}{2\tau_2}|m-\tau \,w|^2 \right] \sum_k \exp[ -4\pi\tau_2\cdot R\,(E'_R)_k +i\cdot 2\pi\tau \cdot P'_k  ] \, .
\end{split}    
\end{equation}
With these, the form of the worldsheet partition function at winding one becomes
\begin{equation}
    \begin{split}
          &\mathcal{Z}_{+}(R,\zeta,\bar{\zeta}) \\= &\; \zeta_2\sum_{m}\sum_w \frac{1}{(\tau_2)_*} \frac{1}{\left( 1-\frac{\mu}{\pi}\frac{r}{R}\,w\right)}\, \mathcal{Z}_{{\rm seed}}^{(m,w)} (\tau_*,\bar{\tau}_*)\\
          =&\; \zeta_2\, \frac{r}{\sqrt{2}} \sum_k  e^{2\pi i\, \tau_* \cdot R(P')_k} \sum_{m}\sum_w \frac{1}{(\tau_2)_*^{3/2}} \frac{1}{\left( 1-\frac{\mu}{\pi}\frac{r}{R}\,w\right)}\, \exp\left[- \frac{\pi r^2}{2(\tau_2)_*}|m-\tau_* \,w|^2 -4\pi (\tau_2)_*\cdot R(E'_R)_k\right] \, .
    \end{split}
\end{equation}
At this localized point, there are simplifications: 
\begin{equation}
    \begin{split}
        m-\bar{\tau}_*\, w=m-\frac{\bar{\zeta}-\frac{\mu}{\pi}\frac{r}{R}m}{1-\frac{\mu}{\pi}\frac{r}{R}w} \,w= \frac{m-\bar{\zeta}\, w}{1-\frac{\mu}{\pi} \frac{r}{R}\, w} \, ,
    \end{split}
\end{equation}
which implies 
\begin{equation}
    \begin{split}
        -\frac{\pi  r^2}{2(\tau_2)_*} (m-\tau_* w)\,(m-\bar{\tau}_*\, w) & = -\frac{\pi r^2}{2} \frac{(1-\frac{\mu}{\pi}\frac{r}{R}w) }{\zeta_2-i\frac{1}{2}\frac{\mu}{\pi}\frac{r}{R}(m-\zeta\, w)} \cdot (m-\zeta\, w)\cdot \frac{m-\bar{\zeta}\, w}{1-\frac{\mu}{\pi} \frac{r}{R}\, w}\\
        &= -\frac{\pi r^2}{2} \frac{|m-\zeta\, w|^2}{\zeta_2-i\frac{1}{2}\frac{\mu}{\pi}\frac{r}{R}(m-\zeta\, w)}\,.
    \end{split}
\end{equation}
As a result, we write the winding one partition function as
\begin{equation}
    \begin{split}
        \mathcal{Z}_{+}(R,\zeta,\bar{\zeta}) = & \zeta_2\, \frac{r}{\sqrt{2}} \sum_k  e^{2\pi i\, \tau_* \cdot R(P')_k} \sum_m \sum_w \frac{(1-\frac{\mu}{\pi}\frac{r}{R}\,w )^{1/2}}{(\zeta_2-i\frac{1}{2}\frac{\mu}{\pi}\frac{r}{R}(m-\zeta\, w))^{3/2}}\\
        \times& \exp\left[  -\frac{\pi r^2}{2} \frac{|m-\zeta\, w|^2}{\zeta_2-i\frac{1}{2}\frac{\mu}{\pi}\frac{r}{R}(m-\zeta\, w)} -4\pi \,\frac{\zeta_2-i\frac{1}{2}\frac{\mu}{\pi}\frac{r}{R}(m-\zeta\, w)}{(1-\frac{\mu}{\pi}\frac{r}{R}w)} R(E'_R)_k\right] \, .
    \end{split}
\end{equation}
To read off the spectrum, we need to translate this to the Hamiltonian picture using Poisson resummation 
\begin{equation}
    \sum_m f\,(m)=\sum_n \widehat{f}\, (n)\,,
\end{equation}
where $\widehat{f}$ is the Fourier transformation of $f$.
Technically, it is easier to do the resummation with respect to  $X$ defined as: 
\begin{equation}
    X= \zeta_2-i\frac{1}{2}\frac{\mu}{\pi}\frac{r}{R}(m-\zeta\, w) \Rightarrow m= \zeta \,w +i\frac{2\pi R}{r\mu}(X-\zeta_2) \, ,
\end{equation}
i.e. $f\,(m)=F(X)$, so that 
\begin{equation}
    \begin{split}
        \widehat{f}(n) & = \int_{-\infty}^{\infty} dm\, f(m)\, e^{-i\cdot 2\pi n\,m}\\
        &= i\frac{2\pi R}{r\mu}\int_{-\infty}^{\infty} dX\, F(X)\, e^{-2\pi i n\cdot (\zeta \,w +i\frac{2\pi R}{r\mu}(X-\zeta_2))}\\
        &= i\frac{2\pi R}{r\mu} e^{-2\pi i n (\zeta \, w -i\frac{2\pi  R}{r\mu}\zeta_2)} \int_{-\infty}^{\infty} dX\, F(X)\,  e^{2\pi n\cdot \frac{2\pi R}{r\mu}X}\,.
    \end{split}
\end{equation}
In our case, we have
\begin{equation}
    \begin{split}
        F(X)& = \frac{1}{(\zeta_2-i\frac{1}{2}\frac{\mu}{\pi}\frac{r}{R}(m-\zeta\, w))^{3/2}} \exp\left[  -\frac{\pi r^2}{2} \frac{|m-\zeta\, w|^2}{\zeta_2-i\frac{1}{2}\frac{\mu}{\pi}\frac{r}{R}(m-\zeta\, w)} -4\pi \,\frac{\zeta_2-i\frac{1}{2}\frac{\mu}{\pi}\frac{r}{R}(m-\zeta\, w)}{(1-\frac{\mu}{\pi}\frac{r}{R}w)} R(E'_R)_k\right]\\
        &=  \frac{1}{X^{3/2}} \exp\left[ -\frac{\pi r^2}{2} \frac{-\frac{4\pi^2 R^2}{r^2\mu^2}(X-\zeta_2)^2+i\frac{4\pi i\,R w \, \zeta_2}{r\mu} (X-\zeta_2) }{X} - 4\pi \frac{R(E'_R)_k }{(1-\frac{\mu}{\pi}\frac{r}{R}w)}\,X\right] \, ,
    \end{split}
\end{equation}
so that the $X$-integral is evaluated as 
\begin{equation}
    \begin{split}
        \int_{-\infty}^{\infty} dX\, F(X)\,   e^{2\pi n\cdot \frac{2\pi R}{r\mu}X} &= \frac{\sqrt{\pi}}{i\pi^{3/2} \frac{\sqrt{2} R\zeta_2}{\mu} \,(1-\frac{\mu}{\pi}\frac{r}{R}w)^{1/2} }\exp\left[ \#\right] \, ,
    \end{split}
\end{equation}
where 
\begin{equation}
    \begin{split}
        \#= -\frac{4\pi^3 R^2 \zeta_2}{\mu^2} \left( 1-\frac{1}{2}\frac{\mu}{\pi}\frac{r}{R}w\right)-i\frac{2\pi^2 R\zeta_2}{\mu} \left[ 8R(E'_R)_k-\frac{4\pi^2 R^2}{\mu^2}\left(1-\frac{\mu}{\pi}\frac{r}{R}w \right) + \frac{8in\,\left(1-\frac{\mu}{\pi}\frac{r}{R}w \right) }{-i\frac{r}{R}\cdot \frac{\mu}{\pi}} \right]^{1/2}\,.
    \end{split}
\end{equation}
Using this, we obtain the final result
\begin{equation}
    \begin{split}
        & \mathcal{Z}_{+}(R,\zeta,\bar{\zeta})\\
        = & \; \zeta_2\, \frac{r}{\sqrt{2}} \sum_k  e^{2\pi i\, \tau_* \cdot R(P')_k} \sum_w \left( 1-\frac{\mu}{\pi}\frac{r}{R}w\right)^{1/2} \sum_n \widehat{f}\,(n)\\
         = & \; \zeta_2\, \frac{r}{\sqrt{2}} \sum_k  e^{2\pi i\, \tau_* \cdot R(P')_k} \sum_w \left( 1-\frac{\mu}{\pi}\frac{r}{R}w\right)^{1/2}  \cdot i\frac{2\pi R}{r\mu} \sum_n e^{-2\pi i n (\zeta \, w -i\frac{2\pi  R}{r\mu}\zeta_2)} \int_{-\infty}^{\infty} dX\, F(X)\,   e^{2\pi n\cdot \frac{2\pi R}{r\mu}X}\\
         = & \; \sum_k \exp\left[ -4\pi\zeta_2 \, R\,(\mathcal{E}_R)_k+2\pi i \zeta\, (R\, P_k)\right]\,,
    \end{split}
\end{equation}
with
\begin{equation}
     R\,(\mathcal{E}_R)_k = \frac{\pi^2}{\mu^2}R\left[ \left( R-\frac{\mu}{\pi}Q\right)-\sqrt{\left( R-\frac{\mu}{\pi}\,Q\right)^2-2\,\frac{\mu^2}{\pi^2} R\,(E_R)_k}\right] \, ,
\end{equation}
where $Q$ is the $U(1)$ charge given by $Q=\frac{n}{r}+\frac{w r}{2}$. Note that the momentum $P_k$ is unchanged.

A similar analysis for $\mathcal{Z}_{-}$ gives the spectrum \eqref{eq:JTbarneg}, which can be interpreted as non-perturbative corrections.
\end{document}